\documentclass[a4paper,12pt]{article}
\usepackage{amsmath}
\usepackage{amsfonts}
\usepackage{amssymb}
\newcommand{\be}{\begin{equation}}
\newcommand{\ee}{\end{equation}}

\newcommand{\ba}{\begin{eqnarray}}
\newcommand{\ea}{\end{eqnarray}}
\usepackage{graphicx}
\def\bfpi{\mbox{\boldmath $\pi$}}
\date{}
\title{Passing the boundary between the parity breaking medium and vacuum by vector particles\footnote{Talk on  IV International conference "Models in Quantum Field Theory" MQFT-2012: 24 - 27 September 2012, Peterhof, Russia  }}
\author{A.A.Andrianov\footnote{ E-mail:  andrianov@icc.ub.edu}, S.S.Kolevatov\footnote{ E-mail:  kss2005@list.ru}\
\\
 Saint Petersburg State University,\\
 198504, Saint Petersburg}
\begin{document}

 \maketitle
\begin{abstract} The electrodynamics supplemented with a Chern-Simons (CS) action (Carrol-Field-Jackiw electrodynamics) in a half space is studied. The passage of photons and massive vector mesons through a boundary between the CS medium and the vacuum of conventional Maxwell electrodynamics is investigated. Effects of reflection from a boundary (up to the total one) are revealed when  vector particles escape  to vacuum and  income from vacuum passing the boundary.
\end{abstract}
\maketitle

\section{Carroll-Field-Jackiw model with the boundary}
 The interest to possible Lorentz and CPT Invariance Violation in the Standard Model was raised up in last 20 years after the seminal paper \cite{Carroll:1989vb}. In that work the  electrodynamics modified with additional Chern-Simons (CS) parity-odd lagrangian spanned on a constant CS four-vector was considered. From the analysis of the radiation of distant radio galaxies it was shown that there is no such violation on the Hubble scales. Nevertheless spontaneous Lorentz symmetry breaking
may occur after condensation of massless axion-like fields \cite{Andrianov:1994qv},\cite{Kostelecky:2002hh}--\cite{ArkaniHamed:2004ar} at  space scales comparable with star and galaxies sizes, in particular, near (or in) very dense stars \cite{mielke}. Axion background can in principle induce the high-energy photon decays into dilepton pairs \cite{axion1} and, in turn,
photon emission by charged particles \cite{axion2}.
Another interesting area for observation of parity breaking is a heavy ion physics. Recently several experiments in heavy ion collisions  have indicated an abnormal yield of lepton pairs \cite{NA60,phenix}. It was conjectured that the effect may be a manifestation of
local parity breaking in colliding nuclei due to generation of pseudoscalar, isosinglet \cite{aaep} or neutral isotriplet \cite{anesp}, classical
background  in finite volume with magnitude depending on the dynamics of the collision.
In the case when axion-like background occurs in  astrophysics or heavy ion physics the existence of a boundary between the parity-odd medium and the vacuum is quite essential. For axion stars there is evidently a boundary where axion background disappears and photons escape  to vacuum. However, not all of the photons penetrate it and partially a reflection arises that will be shown in Sec.2. A similar phenomenon occurs during external irradiation of a star with broken parity and, in addition to, a phase rotation of circularly polarized wave happens.

In heavy ion collisions the yield of photon/vector meson decays into lepton pairs may be enhanced  by pseudoscalar background inside the hot fireball. Such a background is  described in average by axial chemical potential which corresponds to the time-like CS vector, while the boundary of the fireball is spatial. This case is investigated in Sec. 3. In both cases the energy range for vector mesons in medium  is found  in which  the total reflection holds.
Thus the examination of how an axion/pion background in a bounded volume can influence on photons and massive vector mesons propagating through a boundary to vacuum and back may serve for detecting parity odd properties of a medium.

We start the description of Carroll-Field-Jackiw model with the Lagrange density which describes the propagation of a vector
field in the presence of a  pseudoscalar axion-like background,
\begin{eqnarray}
{\mathcal L} &=&  -\,{\textstyle\frac14}\,F^{\alpha\beta}(x)F_{\alpha\beta}(x)
-\,{\textstyle\frac14}\,F^{\mu\nu}(x)\widetilde F_{\mu\nu}(x)\,a_{c\ell}(x)
 + {\textstyle\frac12}\,m^2\,A_\nu(x)A^\nu(x) ,\label{lagrangian1}
\end{eqnarray}
where $A_\mu$ and $a_{c\ell}$ stand for the vector and background  pseudoscalar fields respectively,
$\widetilde F^{\mu\nu}={\textstyle\frac12}\,
\varepsilon^{\,\mu\nu\rho\sigma}\,F_{\,\rho\sigma}$ is a dual field strength. We have included the mass term for vector fields to account for parity breaking effects in the formation of massive vector mesons ($\rho,\omega,\ldots$) in heavy ion collisions\cite{aaep} and to ensure consistency of the photon dynamics  in the case of CS Lagrangian \cite{AACGS2010} with time-like CS vector.

Let us consider a slowly varying classical pseudoscalar background,
\begin{equation}
a_{c\ell}(x)=\,\zeta_\lambda x^{\lambda}\,\theta(-\,\zeta\cdot x)\label{background}
\end{equation}
where $\theta(\cdot)$ is the Heaviside step distribution,
in which a fixed constant four vector $\zeta^\mu$ with mass dimension  is put as an argument.
In this case Lorentz invariance is violated in the Minkowski half space
\mbox{$\zeta\cdot x<0\,.$} Taking in account (\ref{background}), using the Gau\ss\ theorem and integrating by parts, one can derive the field equations from the equivalent Lagrange density,
\begin{eqnarray}
{\mathcal L} =  - \frac14 F^{\alpha\beta}(x)F_{\alpha\beta}(x)
+ \frac12\zeta_\mu A_\nu(x)\widetilde F^{\mu\nu}(x)\,\theta(-\,\zeta\cdot x)
+  \frac{m^2}{2} A_\nu(x)A^\nu(x),\label{lagrangian2}
\end{eqnarray}
The corresponding field equations are,
\begin{eqnarray}
\left\lbrace
\begin{array}{cc}
\Box A^\nu(x)  +  m^2\,A^\nu(x) = \varepsilon^{\,\nu\alpha\rho\sigma}\,\zeta_\alpha\,\partial_\rho A_\sigma(x)
& \qquad{ for}\ \zeta\cdot x<0\\
\Box A^\nu(x)  +  m^2\,A^\nu(x)=0
& \qquad{ for}\ \zeta\cdot x>0\\
\end{array}\right. \label{EL2}
\end{eqnarray}
We shall solve this equations in case of $\zeta\cdot x<0$. Let us recall the construction of the chiral polarization vectors
for the Maxwell-Chern-Simons (MCS) vector field, which are built using the projector on the plane transverse vectors $k_\mu, \zeta_\nu$ \cite{AACGS2010},
\begin{equation}
S^{\nu}_{\phantom{\nu}\lambda}\equiv=
\delta^{\,\nu}_{\;\lambda}\,{\mbox{\tt D}}
+ k^{\nu}\,k_{\lambda}\,\zeta^2 + \zeta^{\,\nu}\,\zeta_{\lambda}\,k^2
- \zeta\cdot k\,(\zeta_{\lambda}\,k^{\nu} + \zeta^{\nu}\,k_{\,\lambda});
\end{equation}
\[
{\mbox{\tt D}}\;\equiv\;(\zeta\cdot k)^2-\zeta^2\,k^2\;=\;\textstyle\frac12\;S^{\nu}_{\phantom{\nu}\nu}.
\]
Using the latter equality one can find that,
\begin{equation}
S^{\,\mu\lambda}\,\varepsilon_{\lambda\nu\alpha\beta}\,\zeta^\alpha k^{\beta}\;
=\;{\mbox{\tt D}}\;\varepsilon^{\,\mu}_{\phantom{\mu}\nu\alpha\beta}\,\zeta^\alpha k^{\beta} .
\end{equation}
Then to our purpose it is convenient to introduce the two orthonormal, one-dimensional, Hermitian projectors,
\begin{equation}
\bfpi^{\,\mu\nu}_{\,\pm}\equiv
\frac{S^{\,\mu\nu}}{2\,{\mbox{\tt D}}}\;
\pm\;\frac{i}{2}\,
\varepsilon^{\mu\nu\alpha\beta}\,\zeta_{\alpha}\,k_{\beta}\,{\mbox{\tt D}}^{\,-\frac12}
=\left(\bfpi^{\,\nu\mu}_{\,\pm}\right)^\ast=\left(\bfpi^{\,\mu\nu}_{\,\mp}\right)^\ast
\qquad\quad(\mbox{\tt D}>0) . \label{projectors1}
\end{equation}
A couple of chiral polarization vectors for the MCS field can be constructed out of constant tetrades  $\epsilon_\nu$,
\begin{eqnarray}
 \varepsilon_{\,\pm}^{\,\mu\ast}(k)\equiv \bfpi^{\,\mu\lambda}_{\,\pm}\,\epsilon_\mu . \label{projectors2}
\end{eqnarray}
Their properties were thoroughly described  in \cite{AACGS2010}.

In order to obtain the normal mode of propagating the MCS field, let us introduce the kinetic $4\times4$ Hermitian kinetic matrix $\mathbb K$ with elements,
\begin{equation}
K_{\,\lambda\nu}\equiv
g_{\,\lambda\nu}\left(k^2-m^2\right) +
i\varepsilon_{\lambda\nu\alpha\beta}\,\zeta^\alpha k^{\beta};\quad
K_{\,\lambda\nu}=K^{\,\ast}_{\,\nu\lambda} .\end{equation}
We obtain the general solution of the free field equations
(\ref{EL2}) for $\zeta\cdot x<0$ from the relations \eqref{projectors1}, \eqref{projectors2},
\begin{eqnarray}
K^{\,\mu}_{\phantom{\mu}\nu}\,\varepsilon^{\,\nu}_{\pm}(k) &=&
\left[\,\delta^{\,\mu}_{\phantom{\mu}\nu}\left(k^2-m^{2}\right)
+ {\sqrt{\mbox{\tt D}}}\,\left(\,\bfpi^{\,\mu}_{\,+\,\nu}\;-\;\bfpi^{\,\mu}_{\,-\,\nu}\,\right)\,\right]
\varepsilon^{\,\nu}_{\pm}(k)\nonumber\\
&=& \left(k^2-m^{2} \pm\,\sqrt{\mbox{\tt D}} \,\right)\,
\varepsilon^{\,\mu}_{\pm}(k) .
\end{eqnarray}
\section{Classical solutions for a space-like Chern-Simons vector}
Let us consider the case of a spatial Chern-Simons vector $\zeta_{\mu}=(0,-\,\zeta_x,0,0)$. In order to solve the classical Euler-Lagrange equations (\ref{EL2}), let us introduce the vectors: $\hat k = (\omega, k_2, k_3)$, $\hat x = (x_0, x_2, x_3) $, and their scalar product $\hat k \cdot \hat x =- \omega x_0 + k_2 x_2 + k_3 x_3$. Using the Fourier transformation in these coordinates one can solve the equation describing $A_1$ in the entire space,
\begin{eqnarray}
\tilde A_1 = \tilde u_{1 \rightarrow}(\omega,k_2,k_3) e^{ik_{10} x_1}+\tilde u_{1 \leftarrow}(\omega,k_2,k_3) e^{-ik_{10} x_1},
\label{A1}
\end{eqnarray}
where
$k_{10}^2 = \omega^2-m^2-k_\bot^2,\ k_\bot^2 = k_2^2 + k_3^2$.\\
Let us consider the system (\ref{EL2}) for the remaining components and perform Fourier transformation over $\hat x$,
\begin{eqnarray}
\left\lbrace
\begin{array}{cc}
(-\omega^2 + m^2 + k_\bot^2)\tilde A_0 - \partial_1^2 \tilde A_0= i \zeta_x \theta(-x_1)(k_2\tilde A_3 - k_3 \tilde A_2)\\
(-\omega^2 + m^2 + k_\bot^2)\tilde A_2 - \partial_1^2 \tilde A_2= -i \zeta_x \theta(-x_1)(k_3\tilde A_0 + \omega \tilde A_3)\\
(-\omega^2 + m^2 + k_\bot^2)\tilde A_3 - \partial_1^2 \tilde A_3= i \zeta_x \theta(-x_1)(\omega \tilde A_2 + k_2 \tilde A_0).\\
\end{array}\right.
\label{systFourier}
\end{eqnarray}
This system is valid in all the space. In the case $x_1>0$ one obtains the solutions of the Maxwell equations in vacuum. But in the case $x_1<0$ solutions describes the MCS electrodynamics,
\ba
\tilde A_\nu = \left\lbrace
\begin{array}{l}
\tilde u_{\nu \rightarrow}(\omega,k_2,k_3)e^{ik_{10} x_1}+\tilde u_{\nu \leftarrow}(\omega,k_2,k_3)e^{-ik_{10} x_1},\!\ x_1>0; \\ \\
\sum\limits_A \left[\tilde v_{\nu A \rightarrow}(\omega,k_2,k_3) e^{ik_{1A} x_1}+\tilde v_{\nu A \leftarrow}(\omega,k_2,k_3) e^{-ik_{1A} x_1}\right], \!\ x_1<0.
\end{array} \right.
\label{anutilde}
\end{eqnarray}
Herein the first index of $\tilde v$ denotes  the corresponding component of $A_\nu$, $\nu=0,2,3$, the second index $A$ stands for different mass-shell dispersion laws $k_1$ for polarizations $L, +, -$ and the arrows $\rightarrow$, $\leftarrow$ point out the direction of particle propagation. The dispersion laws for different polarizations read,
\begin{eqnarray}
\left\lbrace
\begin{array}{l}
k_{1L}=k_{10}=\sqrt{\omega^2-m^2-k_\bot^2};\\
k_{1+}=\sqrt{\omega^2-m^2-k_\bot^2+\zeta_x \sqrt{\omega^2-k_\bot^2}};\\
k_{1-}=\sqrt{\omega^2-m^2-k_\bot^2-\zeta_x \sqrt{\omega^2-k_\bot^2}} .
\end{array}\right.\
\end{eqnarray}
Furthermore, $v$ satisfy the following conditions,
\begin{eqnarray}
\left\lbrace
\begin{array}{ll}
\tilde v_{2+\leftrightarrows}=\frac{k_2k_3-i\omega\sqrt{\omega^2-k_\bot^2}}{\omega^2-k_2^2}\tilde v_{3+\leftrightarrows}; &
\tilde v_{2-\leftrightarrows}=\frac{k_2k_3+i\omega\sqrt{\omega^2-k_\bot^2}}{\omega^2-k_2^2}\tilde v_{3-\leftrightarrows};
\\
\tilde v_{0+\leftrightarrows}=-\frac{i(k_3\tilde v_{2+\leftrightarrows}-k_2\tilde v_{3+\leftrightarrows})}{\sqrt{\omega^2-k_\bot^2}};&
\tilde v_{0-\leftrightarrows}=\frac{i(k_3\tilde v_{2-\leftrightarrows}-k_2\tilde v_{3-\leftrightarrows})}{\sqrt{\omega^2-k_\bot^2}};\\
\tilde v_{2L\leftrightarrows}=\frac{k_2}{k_3}\tilde v_{3L\leftrightarrows}; &
\tilde v_{0L\leftrightarrows}=-\frac{\omega}{k_3}\tilde v_{3L\leftrightarrows}.\\
\end{array}\right.
\label{v}
\end{eqnarray}
 Thus we have the solutions in both half-spaces, and we have to match them on the boundary. Let us integrate the system (\ref{systFourier}) over $x_1$ from $-\varepsilon$ to $\varepsilon$. It will give us \cite{Andrianov:2011wj} the matching conditions,
\begin{eqnarray}
\!\!\!\!
\begin{array}{l}
k_{10}^2(-\frac{\tilde u_{0 \rightarrow}-\tilde u_{0 \leftarrow}}{i k_{10}}+\sum\limits_A \frac{\tilde v_{0A\rightarrow}-\tilde v_{0A\leftarrow}}{i k_{1A}})= -i \zeta_x \sum\limits_A (k_2\frac{\tilde v_{3A\rightarrow}-\tilde v_{3A\leftarrow}}{i k_{1A}}-k_3\frac{\tilde v_{2A\rightarrow}-\tilde v_{2A\leftarrow}}{i k_{1A}});
\\
k_{10}^2(-\frac{\tilde u_{2 \rightarrow}-\tilde u_{2 \leftarrow}}{i k_{10}}+\sum\limits_{A}\frac{\tilde v_{2A\rightarrow}-\tilde v_{2A\leftarrow}}{i k_{1A}})=  i \zeta_x \sum\limits_{A}(k_3\frac{\tilde v_{0A\rightarrow}-\tilde v_{0A\leftarrow}}{i k_{1A}} + \omega\frac{\tilde v_{3A\rightarrow}-\tilde v_{3A\leftarrow}}{i k_{1A}});
\\
k_{10}^2(-\frac{\tilde u_{3 \rightarrow}-\tilde u_{3 \leftarrow}}{i k_{10}}+\sum\limits_{A}\frac{\tilde v_{3A\rightarrow}-\tilde v_{3A\leftarrow}}{i k_{1A}})=- i \zeta_x \sum\limits_{A}(\omega\frac{\tilde v_{2A\rightarrow}-\tilde v_{2A\leftarrow}}{i k_{1A}} + k_2\frac{\tilde v_{0A\rightarrow}-\tilde v_{0A\leftarrow}}{i k_{1A}}) .
\label{bound}
\end{array}
\end{eqnarray}

The contributions to the amplitudes in the right half-space ($u_\mu$) from  different polarizations in the left one are independent: $u_\mu=u_\mu^{(L)}+u_\mu^{(+)}+u_\mu^{(-)}$.
If $A=L$,
the right parts in this system are equal to zero (from \eqref{v}), therefore we get,
\begin{eqnarray}
\tilde u_{\nu \rightarrow}^{(L)}-\tilde u_{\nu \leftarrow}^{(L)}=\tilde v_{\nu L \rightarrow} - \tilde v_{\nu L \leftarrow}
\end{eqnarray}
For $A=\pm$, using the relations between $v_\pm$  (\ref{bound}) one can obtain,
\begin{eqnarray}
\tilde u_{\nu \rightarrow}^{(\pm)}-\tilde u_{\nu \leftarrow}^{(\pm)}=\frac{(\tilde v_{\nu \pm \rightarrow} - \tilde v_{\nu \pm \leftarrow})k_{1+}}{k_{10}} .
\end{eqnarray}
Besides, all contributions in $A$ are continuous,
\begin{eqnarray}
\tilde u_{\nu \rightarrow}^{(A)}+\tilde u_{\nu \leftarrow}^{(A)}=\tilde v_{\nu A \rightarrow} + \tilde v_{\nu A \leftarrow} . \label{cont}
\end{eqnarray}
With the help of this equalities one can obtain the relations between $u$ and $v$ in the final form,
\begin{eqnarray}
\tilde u_{\nu \rightarrow}^{(A)}=\frac{1}{2}(\tilde v_{\nu A \rightarrow}(\frac{k_{1A}+k_{10}}{k_{10}}) - \tilde v_{\nu A \leftarrow}(\frac{k_{1A}-k_{10}}{k_{10}}));\\
\tilde u_{\nu \leftarrow}^{(A)}=\frac{1}{2}(-\tilde v_{\nu A \rightarrow}(\frac{k_{1A}-k_{10}}{k_{10}}) + \tilde v_{\nu A \leftarrow}(\frac{k_{1A}+k_{10}}{k_{10}})).
\label{uv}
\end{eqnarray}

\subsection{Escaping from the parity breaking medium}

 Let the particle pass from the left half-space to the right one. Then one has  to put $\tilde u_{\mu \leftarrow}=0$ and thus,
\begin{eqnarray}
\tilde u_{\nu \rightarrow}^{(L)}=\tilde v_{\nu L \rightarrow} - \tilde v_{\nu L \leftarrow};\quad
\tilde u_{\nu \rightarrow}^{(\pm)}=\frac{(\tilde v_{\nu \pm \rightarrow} - \tilde v_{\nu \pm \leftarrow})k_{1\pm}}{k_{10}},
\end{eqnarray}
for $\nu = 0, 2, 3$ . Using \eqref{cont}, one can find what part is reflected,
\begin{eqnarray}
\tilde v_{\nu L \leftarrow}=0;\quad
\tilde v_{\nu \pm \leftarrow} = \frac{k_{1\pm}-k_{10}}{k_{1\pm}+k_{10}} \tilde v_{\nu \pm \rightarrow} ,
\end{eqnarray}
and what has passed through the boundary,
\begin{eqnarray}
\tilde u_{\nu \rightarrow}^{(L)}=\tilde v_{\nu L \rightarrow};\qquad
\tilde u_{\nu \rightarrow}^{(\pm)}=\frac{2k_{1\pm}}{k_{10}+k_{1\pm}}\tilde v_{\nu \pm \rightarrow} .
\end{eqnarray}
We may rewrite dispersion laws in terms of the invariant mass $M^2=k_{\mu}k^{\mu}=m^2-\zeta_x\sqrt{(\omega^2-k_\bot^2)}$. In the case of the dilepton decay of the particle in parity odd medium, it is an invariant mass of the lepton pair, which is the observable quantity. Now let us express $\omega$ by $M$,
\begin{eqnarray}
k_{1L}=\sqrt{\frac{(M^2-m^2)^2}{\zeta^2}-m^2};\quad
k_{1\pm}=\sqrt{\frac{(M^2-m^2)^2}{\zeta^2}-M^2} .
\end{eqnarray}
In this terms the dispersion laws for $(+)$ and $(-)$ polarizations coincide. But the difference between this expressions is in their domains of definition: when $A=+$, the invariant mass satisfies the condition $M^2<(\sqrt{m^2+\frac{\zeta^2}{4}}-\frac{\zeta}{2})^2$, while for $M^2>(\sqrt{m^2+\frac{\zeta^2}{4}}+\frac{\zeta}{2})^2$  we deal with polarization $(-)$. Notice, that in the case of (+) polarization, the squared invariant mass may be negative. For instance, if the particle is massless, then for the positive polarization of the photon one has $M^2<0$. Nevertheless, on the whole set of accessible values of the invariant mass, we may write a reflection coefficient in the form,
\be
k_{ref}=\frac{|\sqrt{\frac{(M^2-m^2)^2}{\zeta^2}-M^2}-
\sqrt{\frac{(M^2-m^2)^2}{\zeta^2}-m^2}|}{|\sqrt{\frac{(M^2-m^2)^2}{\zeta^2}-M^2}
+\sqrt{\frac{(M^2-m^2)^2}{\zeta^2}-m^2}|}.
\ee
Such a dependence on the invariant mass allows us to plot the $k_{ref}(M^2)$. We present here the graph for the photon escaping from the parity breaking region (fig. 1).
\begin{figure}[h]
\center{\includegraphics[width=0.9\linewidth]{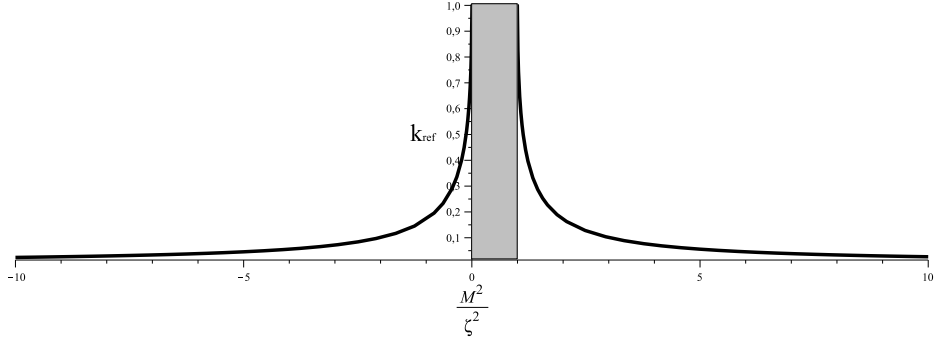}}
\caption{The reflection coefficient  from the boundary for  photons escaping the parity odd region. The kinematically forbidden range of values of the invariant mass is shaded.} 
\end{figure}\\
As one can see from the graph, in the case of the spatial CS vector and the spatial boundary, for $|M^2|>>\zeta^2$ most photons escape from the parity odd medium, while a significant part of the photons with $M^2\sim\zeta^2$ for (-) polarization and with $|M^2|<<\zeta^2$ for (+) polarization reflects from the boundary and does not escape from the medium.

\subsection {Entrance to the parity odd medium}
It is no less interesting to consider another case, when a particle moves from the right half-space to the left one.
Respectively, we have to take $\tilde v_{\mu A \rightarrow}=0$. From relations (\ref{uv}) we obtain,
\begin{eqnarray}
\tilde u_{\nu \rightarrow}^{(A)}=\frac{k_{10}-k_{1A}}{k_{10}+k_{1A}} \tilde u_{\nu \leftarrow}^{(A)};\quad
\tilde v_{\nu A \leftarrow}=\frac{2k_{10}}{k_{10}+k_{1A}}\tilde u_{\nu \leftarrow}^{(A)} \label{vA}
\end{eqnarray}
The only point we do not know in this problem is the relation between $\tilde u_{\nu \leftarrow}^{(+)},\tilde u_{\nu \leftarrow}^{(-)}, \tilde u_{\nu \leftarrow}^{(L)}$. We can find them using (\ref{v}), (\ref{vA}) and all incoming amplitudes $(\tilde u_{\nu \leftarrow})$.
Below the expressions for all components are presented,
\begin{eqnarray}
\left\lbrace
\begin{array}{l}
\tilde u_{0 \leftarrow}^{(L)}=\frac{\omega^2}{\omega^2-k_{\bot}^2}\tilde u_{0 \leftarrow}+\frac{\omega k_3}{\omega^2-k_{\bot}^2}\tilde u_{3 \leftarrow} + \frac{\omega k_2}{\omega^2-k_{\bot}^2}\tilde u_{2 \leftarrow};\\
\tilde u_{0 \leftarrow}^{(\pm)}=-\frac{k_{\bot}^2}{2(\omega^2-k_{\bot}^2)}\tilde u_{0 \leftarrow}-\frac{\omega k_3\mp i k_2 \sqrt{\omega^2 - k_{\bot}^2}}{2(\omega^2-k_{\bot}^2)}\tilde u_{3 \leftarrow}-\frac{\omega k_2\pm i k_3 \sqrt{\omega^2 - k_{\bot}^2}}{2(\omega^2-k_{\bot}^2)}\tilde u_{2 \leftarrow}.
\label{u0}
\end{array}\right.
\end{eqnarray}
\begin{eqnarray}
\left\lbrace
\begin{array}{l}
\tilde u_{2 \leftarrow}^{(L)}=-\frac{k_2^2}{\omega^2-k_{\bot}^2}\tilde u_{2 \leftarrow}-\frac{\omega k_2}{\omega^2-k_{\bot}^2}\tilde u_{0 \leftarrow} - \frac{k_2 k_3}{\omega^2-k_{\bot}^2}\tilde u_{3 \leftarrow};\\
\tilde u_{2 \leftarrow}^{(\pm)}=\frac{\omega^2 - k_3^2}{2(\omega^2-k_{\bot}^2)}\tilde u_{2 \leftarrow}+\frac{\omega k_2\mp i k_3 \sqrt{\omega^2 - k_{\bot}^2}}{2(\omega^2-k_{\bot}^2)}\tilde u_{0 \leftarrow}+\frac{k_2 k_3\mp i \omega \sqrt{\omega^2 - k_{\bot}^2}}{2(\omega^2-k_{\bot}^2)}\tilde u_{3 \leftarrow}.
\end{array}\right.
\end{eqnarray}
\begin{eqnarray}
\left\lbrace
\begin{array}{l}
\tilde u_{3 \leftarrow}^{(L)}=-\frac{k_3^2}{\omega^2-k_{\bot}^2}\tilde u_{3 \leftarrow}-\frac{\omega k_3}{\omega^2-k_{\bot}^2}\tilde u_{0 \leftarrow} - \frac{k_2 k_3}{\omega^2-k_{\bot}^2}\tilde u_{2 \leftarrow};\\
\tilde u_{3 \leftarrow}^{(\pm)}=\frac{\omega^2 - k_2^2}{2(\omega^2-k_{\bot}^2)}\tilde u_{3 \leftarrow}+\frac{\omega k_3\pm i k_2 \sqrt{\omega^2 - k_{\bot}^2}}{2(\omega^2-k_{\bot}^2)}\tilde u_{0 \leftarrow}+\frac{k_2 k_3\pm i \omega \sqrt{\omega^2 - k_{\bot}^2}}{2(\omega^2-k_{\bot}^2)}\tilde u_{2 \leftarrow} .
\label{u3}
\end{array}\right.
\end{eqnarray}
Now with the help of these relations, we can find what part of photons is reflected and what is passed through. Photon can have two transversal polarizations. We may obtain the solution in the right half-space ($x_1>0$), using the variables $\hat k$ и $\hat x$,
\ba
\label{clsol}
A^{\mu}(x) &=& \int\mathrm d\hat k \,\theta(\omega^2 - k^2_\bot )\sum_{r=1}^2\
\left[\,a_{\,{\hat k}\,,\,r}\,u^{\,\mu}_{\,{\hat k}\,,\,r}(x)
+ a^\ast_{\,{\hat k}\,,\,r}\,
u^{\,\mu\,\ast}_{\,{\hat k}\,,\,r}(x)\,\right],\\
u^\nu_{{\hat k}\,,\,r}(x)&=&[\,(2\pi)^3\,2k_{10}\,]^{-1/2}\,
e_{\,r}^{\,\nu}({\hat k})\,
\exp\{\,i\,k_{10} x_1 + i\,{\hat k}\cdot{\hat x}\},\quad r=1,2.
\ea
The two linear transversal polarization vectors are orthonormal and
on the mass shell $k^2=0\,$ do satisfy the closure relations,
\ba
k^\mu e^i_\mu(\hat k)=0;\quad
g^{\mu \nu} e^i_\mu(\hat k) e^{j}_{\nu}(\hat k)=\delta ^{ij}. \label{polar}
\ea
Without loss of generality, let us fix the polarizations,
\ba
e^1_\mu(\hat k)=(0,0,\frac{k_3}{|k_\bot|},-\frac{k_2}{|k_\bot|});\quad
e^2_\mu(\hat k)=\frac{i}{|k_\bot|\sqrt{\omega^2-k_\bot^2}}(k_\bot^2,0,\omega k_2, \omega k_3).
\ea
We may consider a situation, when a ray of polarized photons with the polarization $e^1(\hat k)$ falls across the boundary from vacuum,
\ba
 \hat u_\leftarrow (\hat k)=a(\hat k) |k_\bot| \hat e^1(\hat k).
\ea
By means of (\ref{u0})-(\ref{u3}), we obtain the reflected part,
\ba
\!\!\!\!\!\!&&\hat u_\rightarrow(\hat k)
=a(\hat k)\frac{|k_\bot|}{2}\\\!\!\!\!\!\!&& \times\left[ (\frac{k_{10}-k_{1-}}{k_{10}+k_{1-}}+\frac{k_{10}-k_{1+}}{k_{10}+k_{1+}}) \hat e^1(\hat k) +  (\frac{k_{10}-k_{1-}}{k_{10}+k_{1-}}-\frac{k_{10}-k_{1+}}{k_{10}+k_{1+}}) \hat e^2(\hat k) \right].\nonumber
\ea
Thereby in this case we may define the two reflection coefficients,
\ba
k_{ref}^{e^1\rightarrow e^1}=\left|\frac12 (\frac{1-\sqrt{1-\frac{\zeta}{k_1}}}{1+\sqrt{1-\frac{\zeta}{k_1}}}
+\frac{1-\sqrt{1+\frac{\zeta}{k_1}}}{1+\sqrt{1+\frac{\zeta}{k_1}}})\right|;\\
k_{ref}^{e^1\rightarrow e^2}=\left|\frac12 (\frac{1-\sqrt{1-\frac{\zeta}{k_1}}}{1+\sqrt{1-\frac{\zeta}{k_1}}}
-\frac{1-\sqrt{1+\frac{\zeta}{k_1}}}{1+\sqrt{1+\frac{\zeta}{k_1}}})\right|;
\ea
using the expression for $k_{1+},k_{1-}$ and taking in account the fact that photons coming from vacuum are on the mass shell ($k_1=k_{10}$).
\begin{figure}[h]
\vspace{-0.5cm}
\center{\includegraphics[width=0.6\linewidth]{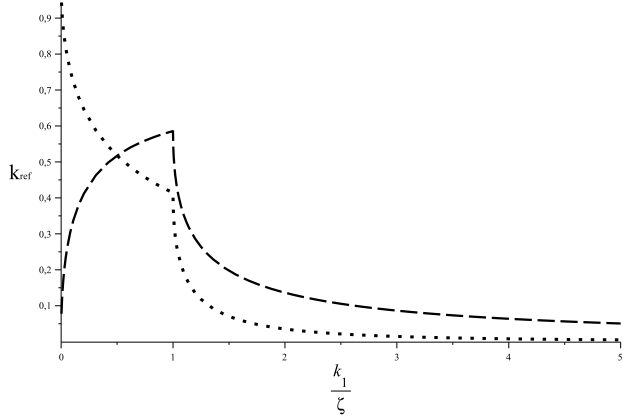}}
\vspace{-0.5cm}
\caption{Reflection of the photons with linear polarization. Dotted line corresponds to the reflection coefficient for the process without polarization change. Dash line is for the process with the change of polarization.}
\end{figure}

At values $k_1 < \zeta$ photons cannot penetrate into the parity odd medium (this process is kinematically forbidden). However, starting from $k_1=\zeta$ photons may cross the boundary, that is why in Fig. 2 one can see a cusp of the curves at this value. Thus, low-energy photons are completely reflected from the boundary; with increasing energy some of photons may change their polarizations; and only when $k_1\geq\zeta$ a part of photons may enter to the CS medium.
Let us construct those polarizations (linear combinations of $e^1$ and $e^2$) which are not mixed in the process of reflection,
\ba
e^L=\frac{1}{\sqrt{2}}(e^1+e^2);\quad
e^R=\frac{1}{\sqrt{2}}(e^1-e^2).
\ea
For this circular polarizations we may define the reflection coefficients,
\ba
k^L_{ref}=\left|\frac{1-\sqrt{1-\frac{\zeta}{k_1}}}{1+\sqrt{1-\frac{\zeta}{k_1}}}\right|;\quad
k^R_{ref}=\left|\frac{1-\sqrt{1+\frac{\zeta}{k_1}}}{1+\sqrt{1+\frac{\zeta}{k_1}}}\right|.
\ea
One should also pay attention to the fact that in the process of reflection circular polarizations change their phases since,
\ba
e^{L} \rightarrow \frac{1-\sqrt{1-\frac{\zeta}{k_1}}}{1+\sqrt{1-\frac{\zeta}{k_1}}} e^{L};\quad
e^{R} \rightarrow \frac{1-\sqrt{1+\frac{\zeta}{k_1}}}{1+\sqrt{1+\frac{\zeta}{k_1}}} e^{R}.
\ea
If $k_1<\zeta$, then after reflection the left polarization changes its phase by $Arg( \frac{1-\sqrt{1-\frac{\zeta}{k_1}}}{1+\sqrt{1-\frac{\zeta}{k_1}}})$. The coefficient for $e^{R}$ on the entire domain \mbox{$k_1>0$} is negative, so the phase change in this case is $\pi$.
\section{Time-like Chern-Simons vector with a spatial boundary}
The case of time-like CS vector and spatial boundary $\zeta_{\mu}=(\,\zeta \theta(-x_1),0,0,0)$ may be useful for a description of processes occurring in heavy ion collisions because it helps to understand what happens with  particles inside the fireball. Just as in the previous section which deals with spatial CS vector one can find solutions for the vector field $А$ in a form (\ref{A1}),(\ref{anutilde}). But in this case the dispersion laws change,
\ba
&&k_{1L}=k_{10}=\sqrt{\omega^2-m^2-k_\bot^2};\nonumber\\
&&k_{1\pm}=\sqrt{\omega^2-m^2-k_\bot^2+\frac{\zeta^2}{2} \mp\zeta \sqrt{\omega^2-m^2+\frac{\zeta^2}{4}}}.
\ea
 As well as for spatial CS vector we may obtain the relations between the coefficients  $\tilde v_{iA\rightleftarrows}$;
\ba
\tilde v_{2A\rightleftarrows}=\frac{-i\zeta k_{1A}(k_{1A}^2-k_{10}^2)-\zeta^2 k_2 k_3}{(k_{1A}^2-k_{10}^2)^2-\zeta^2 k_3^2}\tilde v_{3A\rightleftarrows}\equiv C_{2A}\tilde v_{3A\rightleftarrows}; \nonumber
\\
\tilde v_{1A\rightleftarrows}=\frac{-i\zeta k_{2}(k_{1A}^2-k_{10}^2)-\zeta^2 k_3 k_{1A}}{(k_{1A}^2-k_{10}^2)^2-\zeta^2 k_3^2}\tilde v_{3A\rightleftarrows}.
\label{rel}
\ea
Now let us consider the system (\ref{EL2}) and perform the Fourier transformation over variables $x_0$, $x_2$, $x_3$.
Just as before, matching conditions which are obtained from the integrating the system over an infinitesimal interval $(-\varepsilon,\varepsilon)$ dictate the continuity of all spatial components of the vector field. Using this fact we can write the reflection coefficients for each polarization for the particles escaping from parity breaking medium.
\be
k_{ref}=|\frac{k_{1A}-k_{10}}{k_{1A}+k_{10}}| .
\label{refl}
\ee
We may plot the dependence of reflection coefficient on the transversal momentum ($k_\bot$) and the invariant mass $M^2=k_{\mu}k^{\mu}$. Let us represent dispersion laws in terms of invariant mass,
\ba
k_{1A}^2=\frac{(M^2-m^2)^2}{\zeta^2}-k_\bot^2;\quad
k_{10}^2=\frac{(M^2-m^2)^2}{\zeta^2}+(M^2-m^2)-k_\bot^2 .
\ea
The dispersion relations in these variables are the same for both polarizations again, however their  domains are different. The change of polarizations occurs at the value $M_0^2=m^2-\frac{\zeta^2}{4}$. It means that if $M<M_0$ then the polarization is negative, but if $M>M_0$ then we deal with the positive polarization. Furthermore there is a condition which arises from the kinematic reasons: $|k_\bot|\leq\frac{|M^2-m^2|}{\zeta}$. Using these arguments one can write the reflection coefficient in a following form,
\be
k_{ref}=\frac{|\sqrt{\frac{(M^2-m^2)^2}{\zeta^2}-k_\bot^2}
-\sqrt{\frac{(M^2-m^2)^2}{\zeta^2}+(M^2-m^2)-k_\bot^2}|}{|\sqrt{\frac{(M^2-m^2)^2}{\zeta^2}
-k_\bot^2}+\sqrt{\frac{(M^2-m^2)^2}{\zeta^2}+(M^2-m^2)-k_\bot^2}|} .
\label{kref}
\ee
Generally speaking $k_{ref}$ depends on $M$ and $k_\bot$. But we present here the graph for $\omega$-meson, which flies by perpendicularly to the boundary, i.e. for this particle $k_\bot=0$. In this case one can see that for the values of  invariant mass $m^2-\zeta^2<M^2<m^2$ the reflection coefficient equals unity.
\begin{figure}[h]
\center{\includegraphics[width=0.6\linewidth]{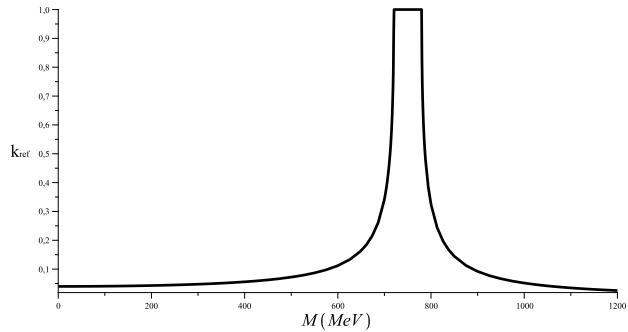}}
\caption{Reflection coefficient for vector meson flying by perpendicularly to the boundary. For vector mesons  $\zeta=300MeV$ is taken.}
\end{figure}
As clearly seen from the graph, there is an interval of invariant masses $(~720-780MeV)$, where a vector meson born in the fireball cannot leave it, because it undergoes a total reflection at these values.

However the choice of CS vector discussed above  ($\zeta_{\mu}=(\,\zeta\theta(-x_1),0,0,0)$) does not provide the gauge invariance since under the gauge transformations the following functional is not conserved,
\ba
\int d^3 x \zeta_\mu A_\nu \partial_\rho A_\sigma \varepsilon^{\mu \nu \rho \sigma},
\label{gauge}
\ea
The gauge invariance can be saved by adding one more component to the CS vector,
\ba
\zeta_\mu=(\zeta \theta (-x_1), \zeta_x(x_1,t), 0, 0);\quad \partial_1 \zeta_0=\partial_0 \zeta_1, \label{CSmod}
\ea
The simplest solution of this differential equation is, $\zeta_x=-\zeta t \delta (x_1)$. Obviously inside the parity breaking medium the solutions of the equations of motion remain the same.

Let us consider the system (\ref{EL2}) with the modified CS vector \eqref{CSmod} and perform the Fourier transformation over variables $x_0$, $x_2$, $x_3$.
\begin{eqnarray}
&&(-\omega^2 + m^2 + k_\bot^2)\tilde A_0 - \partial_1^2 \tilde A_0= - \zeta \partial_{\omega}(k_3\tilde A_2 - k_2 \tilde A_3) \delta(x_1);\nonumber\\
&&(-\omega^2 + m^2 + k_\bot^2)\tilde A_1 - \partial_1^2 \tilde A_1= - i \zeta \theta(-x_1) (k_3\tilde A_2 - k_2 \tilde A_3);\\
&&(-\omega^2 + m^2 + k_\bot^2)\tilde A_2 - \partial_1^2 \tilde A_2= \nonumber\\&&\phantom{(-\omega^2 + m^2 + k_\bot^2)}=\zeta \theta(-x_1)(i k_3\tilde A_1 + \partial_1 \tilde A_3) - \zeta \partial_{\omega}(k_3\tilde A_0 + \omega \tilde A_3)\delta(x_1);\nonumber\\
&&(-\omega^2 + m^2 + k_\bot^2)\tilde A_3 - \partial_1^2 \tilde A_3= \nonumber\\&&\phantom{(-\omega^2 + m^2 + k_\bot^2)}=\zeta \theta(-x_1)(- i k_2\tilde A_1 - \partial_1 \tilde A_2) + \zeta \partial_{\omega}(k_2\tilde A_0 + \omega \tilde A_2)\delta(x_1).\nonumber
\label{systFourierTC}
\end{eqnarray}
When integrating this system over $x_1$ from $-\varepsilon$ to $\varepsilon$ we find new matching conditions,
\ba
\left\lbrace
\begin{array}{l}
i k_{10}(\tilde u_{0\leftarrow}^{(A)}-\tilde u_{0\rightarrow}^{(A)}+\tilde u_{0\rightarrow}^{(A)}-\tilde u_{0\leftarrow}^{(A)})=-\zeta \partial_{\omega}((k_3C_{2A}-k_2)(\tilde v_{3A\rightarrow}+\tilde v_{3A\leftarrow}));\\
\partial_1 \tilde A_1 - \text{continuous};\\
\partial_{\omega}(\tilde u_{0\leftarrow}^{(A)}+\tilde u_{0\rightarrow}^{(A)})=-\frac{\omega \partial_{\omega}C_{2A}}{k_2-k_3 C_{2A}}(\tilde v_{3A\rightarrow}+\tilde v_{3A\leftarrow});\\
\partial_{\omega}(\tilde v_{3A\rightarrow}+\tilde v_{3A\leftarrow})=(-\frac{1}{\omega}+\frac{k_3 \partial_{\omega}C_{2A}}{k_2-k_3 C_{2A}})(\tilde v_{3A\rightarrow}+\tilde v_{3A\leftarrow}).
\end{array}\right.
\ea
The second of these conditions and Eq.(\ref{rel}) provide the continuity of all spatial components of the vector potential. Taking this into account we may use the results obtained above. Thus the expression for the reflection coefficient (\ref{kref}) remains the same and effects connected with  the total (or partial) reflection appear in the same form.
\section{Conclusions}
In this paper we studied the model of physical phenomena related to the propagation of photons and vector mesons between parity breaking medium and vacuum. Main results were obtained for the spatial CS vector.  The relations are presented which are suitable to calculate  the passage or reflection of incoming or outgoing particles from the domain for each polarization. In particular it was shown that transversal polarizations undergo strong reflection (up to the total internal one) at certain values of frequency. The analogous relations are found for the time-like CS vector but with a spatial boundary when there are problems with gauge invariance. It was shown that the mentioned problems do not affect the reflection and transmission coefficients which  are just observable quantities in this case. In addition, for the spatial CS vector it was revealed that during the irradiation of parity odd medium by photons  an additional rotation of circular polarizations may take place upon reflection from the boundary.
Thereby the influence of a boundary between parity breaking medium and vacuum on the decay width of photons and vector mesons represents an important and interesting problem which deserves to be a subject of further investigation.

{\bf Acknowledgments. } This work is supported by Grant RFBR 13-02-00127. S.S. Kolevatov acknowledges the financial support from the non-profit Dynasty Foundation.

\end{document}